\documentclass[prl,twocolumn,floatfix]{revtex4}
\usepackage{dcolumn,amsmath}
\usepackage{bm}

\def\etal{{\it et al.}\ }

\newcommand{\Eref}[1]{Eq.~(\ref{#1})}
\newcommand{\tref}[1]{Table~\ref{#1}}

\begin{document}
\title{In search of the electron electric dipole moment: relativistic
        correlation calculations of the P,T-violation effect
         in the ground state of HI$^+$}.
\author{T.A.\ Isaev}\email{timisaev@pnpi.spb.ru}
\author{A.N.\ Petrov}
\author{N.S.\ Mosyagin}
\author{A.V.\ Titov}
  \email{titov@pnpi.spb.ru}
\affiliation{Petersburg Nuclear Physics Institute, Gatchina, 188300, Russia}

\begin{abstract}
 We report the first results of {\em ab initio} relativistic correlation
 calculation of the effective electric field on the electron, $E_{\rm eff}$, in
 the ground state of the HI$^+$ cation.  This value is required for
 interpretation of the suggested experiment on search for the electron electric
 dipole moment.  The generalized relativistic effective core potential,
 Fock-space relativistic coupled cluster with single and double cluster
 amplitudes and spin-orbit direct configuration interaction methods are used,
 followed by nonvariational one-center restoration of the four-component
 wavefunction in the iodine core.  The calculated value of $E_{\rm eff}$ by the
 coupled cluster method is $E_{\rm eff}=0.345\times 10^{24}$Hz/$e \cdot$cm.
 Configuration interaction study gives $E_{\rm eff}=0.336\times 10^{24}$Hz/$e
 \cdot$cm (our final value).
 The structure of chemical bonding and contributions to $E_{\rm eff}$ in HI$^+$
 is clarified and significant deviation of our value from that obtained in
 Ravaine \etal Phys.\ Rev.\ Lett.\ {\bf 94}, 013001 (2005) is explained.
\end{abstract}

\maketitle

\paragraph*{Introduction.}

 It is known \cite{Khriplovich:97,Commins:99} that existence of the permanent
 electric dipole moments (EDM) of the elementary particles violate two
 fundamental symmetries: space parity (P) and time reversal (T). Considerable
 experimental efforts invested recently to the search for the electron EDM
 $d_e$ (see \cite{Romalis:01a,Regan:02, Hudson:02,Sauer:05Aa, DeMille:00}) are
 primarily connected with the high sensitivity of $d_e$ to the ``new physics''
 beyond the Standard Model (see \cite{Commins:99} and references).  Polar
 heavy-atom diatomics with nonzero projection of the total electronic momentum
 on the molecular axis (see below) are currently considered as the most
 prospective objects to search for $d_e$ because of the great value of the
 effective electric field acting on the unpaired electrons in the ground or
 excited states of such molecules \cite{Sushkov:78, Gorshkov:79}.  The only
 nonzero component of the effective electric field in polar diatomics is the
 one directed along the molecular axis and is traditionally written as $E_{\rm
 eff} \equiv W_d |\Omega|$ where $\Omega$ is the projection of total electron
 angular momentum $\bm{J}$ to the molecular axis.  (see \Eref{WdOmega} and
 \cite{Kozlov:95, Titov:05b} for details).  Calculation of the $E_{\rm eff}$
 value is needed for any experiment on the $d_e$ search using molecules.  The
 theoretical methods developed recently (see \cite{Titov:05a, Titov:05b} and
 references therein) allow one to calculate $E_{\rm eff}$ for any diatomic
 molecule of practical interest with required accuracy, even for such
 complicated system as excited states in PbO
  \cite{Isaev:04, Petrov:05a}.

 Recently, the EDM experiment of new type, on the molecular HI$^+$ cation in a
 trap, was suggested by Stutz and Cornell \cite{Stutz:04}.  The estimates for
 the value of $E_{\rm eff}$ in two molecular cations, HBr$^+$ and HI$^+$, were
 made by Ravaine \etal \cite{Ravaine:05a} and markable influence of the
 chemical bond nature was emphasized.  In the present article, we report the
 first results of {\it ab initio} calculation of $E_{\rm eff}$ for the ground
 state of HI$^+$, leaving out discussion about sensitivity of the suggested
 experiment.

 Following \cite{Chanda:95}, Ravaine \etal modelled electronic structure of
 HI$^+$ in \cite{Ravaine:05a} by two limiting approximations:  ``ionic'' and
 ``covalent'', where ``ionic'' approximation corresponds to a neutral iodine
 atom electrostatically perturbed by a proton.  The latter is located at the
 experimental equilibrium distance from the iodine nucleus determined for the
 HI$^+$ ground state.  The ``covalent'' limit corresponds to the I$^+$ ion
 perturbed by the electrostatic field from the dipole of the polarized neutral
 H.  We are using inverted commas to mark the approximations since from the
 traditional point of view they both correspond rather to a
 van\,der\,Waals--type interaction between ion and induced dipole.
 It was found in \cite{Ravaine:05a} that the value of $E_{\rm eff}$ is greatly
 changed (about six times) depending on the approximation made, either
 ``ionic'' or ``covalent''.

\paragraph*{Model consideration.}

 We calculated molecular dipole moment $D$ of the ground HI$^+$ state with the
 molecular axes origin at the iodine nucleus using the restricted active space
 self-consistent field (RASSCF) \cite{MOLCAS} method.  The details of that
 calculation can be found on \cite{httpI}.
 Our calculation shows that the highest doubly occupied $\sigma$-orbital is
 bonding and most ``mixed'' one among occupied orbitals.  It is formed mainly
 by the iodine $5p_0$ and hydrogen $1s$ orbitals, where subscript denotes the
 projection of the angular momentum on the molecular axis.  Though this is not
 the highest by energy from the occupied orbitals, it gives 77\% of the
 calculated electronic part of
 $D$, $D_{el}=-2.610$ (the contribution from the hydrogen nucleus is,
 obviously, $1{\cdot}R_e=3.08$; here and below we use atomic units unless the
 opposite is stated explicitly).  On the other hand, the valence $\pi_{-1}$ and
 $\pi_{+1}$ orbitals are formed mainly by the iodine $5p_{-1}$ and $5p_{+1}$
 orbitals (mixture of the $5p_{1/2,\pm1/2}$, $5p_{3/2,\pm1/2}$, and
 $5p_{3/2,\pm3/2}$ spinors) and their contribution to $D_{el}$ is about 7\%.

    The distinctions of the ``covalent'' and ``ionic'' approximations from our
    consideration
    can be illustrated on
    the one-configurational model of
  the chemical bond in HI$^+$. The leading (SCF) configuration of this
  molecule in the ground $^2\Pi_{3/2}$ state with $\Omega=+3/2$
  (having weight $0.9$ in the correlated wavefunction)
  can be presented as
  $[\dots]\sigma^2(\pi_{-1,\alpha}\pi_{+1,\beta})\pi'_{+1,\alpha}$, where
  $\alpha$ and $\beta$ correspond to the spin projections $+1/2$ and $-1/2$,
  the degenerate $\pi_{-1,\alpha}$ and $\pi_{+1,\beta}$
  spin-orbitals constitute a shell in the used relativistic classification, and
  the unpaired $\pi'_{+1,\alpha}$ state is mainly the $5p_{3/2,+3/2}$ spinor of
  iodine (we distinguish $\pi'$ from $\pi$ just to emphasize their different
  behavior at the iodine nucleus).
%
  Let us consider contributions of the bonding spin-orbitals
  $|\sigma_{\alpha,\beta}\rangle =
   C_I|\sigma^I_{\alpha,\beta}\rangle + C_H|\sigma^H_{\alpha,\beta}\rangle$
  to the hyperfine structure (HFS) properties at the iodine nucleus.
  The occupied $\sigma^2$ shell can be presented as
\begin{eqnarray}
 \label{sigma^2}
    \sigma^2 \equiv \sigma_{\alpha} \sigma_{\beta}
 &=& C_I^2\ \sigma^I_{\alpha} \sigma^I_{\beta}\nonumber\\
 &+& C_I C_H \left( \sigma^I_{\alpha} \sigma^H_{\beta}
                   + \sigma^H_{\alpha} \sigma^I_{\beta} \right)
 \label{sigma_I-sigma_H}\\
 &+& C_H^2\ \sigma^H_{\alpha} \sigma^H_{\beta}\nonumber\ ,
\end{eqnarray}
  where the operator of asymmetrization is omitted,
  $|\sigma^I\rangle$ is mainly $5p_0$-orbital of iodine,
  $|\sigma^H\rangle$ is mainly $1s$-orbital of hydrogen, and $C_I^2+C_H^2=1$
  (assume for simplicity that $|\sigma^I\rangle$ and $|\sigma^H\rangle$
  are orthogonal and $C_I,C_H$ are real;
  $C_I{\approx}0.83, C_H{\approx}0.56$ in our calculations).
  In the spin-orbit representation, the atomic $5p_0$ orbital can be
  approximately presented as a combination of the $5p_{1/2}$ and $5p_{3/2}$
  spinors of iodine with weights $1/3$ and $2/3$, respectively.

  The contribution in the first line of \Eref{sigma^2} corresponds to the
  ``ionic'' model in \cite{Chanda:95,Ravaine:05a} (when $|C_I|\sim1$), two terms
  in the second line can be compared to their ``covalent'' model (the maximum
  $|C_IC_H|{=}1/2$ is, obviously, attained for $|C_I|{=}|C_H|=1/\sqrt{2}$) and
  the term in the third line corresponds to the conventional ionic model
  \mbox{I$^{++}$--H$^-$} ($|C_H|\sim1$) that is not considered there.
  Both terms from the second line equally contribute to the electric quadrupole
  HFS constant on iodine
  (dependent on the space-inhomogeneous part of electronic density with respect
  to the iodine nucleus) whereas their contributions are completely compensated
  for the spin-dependent magnetic dipole HFS constants as well as for
  $E_{\rm eff}$ (see below).  The latter is a consequence of the fact that
  closed shells do not influence on those spin-dependent properties; their
  contributions can become nonzero only when polarization and
  correlation effects are considered.  So, the only open shell
  $\pi'_{+1,\alpha}$ ($5p_{3/2,+3/2}$)
  should be considered when calculating spin-dependent properties within our
  simple one-configuration model.  The weight of the rest configurations
  (obviously, orthogonal to the leading one and accounting for correlation
  without any restriction on the occupancy of
  $\pi_{\pm1,\alpha},
  \pi_{\pm1,\beta}, \sigma^{I,H}_{\alpha}, \sigma^{I,H}_{\beta}$ etc.) is only
  $0.1$ (as is obtained in the
  calculations discussed below).
    Even if one suggests that all the correlating configurations contain
    singly-occupied $1s$-orbital of hydrogen and $\sigma^H$ is also $1s$, the
    maximal weight of the HI$^+$ configurations of type $[\dots]_I 1s^1_H$ is
    smaller than $0.7$ that can be compared to the weight $1.0$ in the
    ``covalent'' model of Chanda \etal \cite{Chanda:95} and Ravaine \etal
    \cite{Ravaine:05a}.
%
%
   Moreover, in the ``covalent'' model by Ravaine \etal the contributions with
   the weights $2/3$ and $1/3$ (see Eq.~(13) in \cite{Ravaine:05a}) in the front
   of the wavefunction terms containing $\sigma^H_{\alpha}$ and
   $\sigma^H_{\beta}$ ($1s$ of hydrogen) are fixed in accord to the
   spin-coupling rules (with the lowest lying $^3\Pi_{2,1}$ states of I$^+$ and
   proper dissociation limit) and not varied.  This induces a large artificial
   asymmetry in contributions of the $5p_{1/2,\pm1/2}$ states when calculating
   $E_{\rm eff}$ within the ``covalent'' model both at the
   one-configuration and correlation levels.
   Such asymmetry could be attained in the correlation calculations only if
   the configuration in which the singlet $\sigma^2$ pair replaced in the
   leading configuration by the triplet $\sigma^I_{\alpha}\sigma^H_{\beta}$
   state would have weight $\approx 0.1$, thus leaving nothing to other
   correlation configurations.
%
%

\paragraph*{Effective P,T-odd Hamiltonian.}

 The terms of our interest for HI$^+$ in the effective spin-rotational
 Hamiltonian
 may be written following Refs.\ \cite{Dmitriev:92,Kozlov:95}.  The P,T-odd
 interaction of $d_e$ with the effective electric field is
%
\begin{equation}
 \label{eq:hd}
   H_d = W_d~d_e (\bm{J} \cdot \bm{n})\ ,
\end{equation}
 where $\bm{J}$ is the total electronic momentum
 and $\bm{n}$ is the unit vector along the molecular axis from I to H.
 In \cite{Ravaine:05a} slightly different form of $H_d$ is used:
\begin{equation}
 \label{eq:hd_D}
   H_d = W_d (\bm{J} \cdot \bm{n})\ ,
\end{equation}
 so the value of $d_e$ appears explicitly in their final result for $E_{\rm
 eff}$.
%
%
 The effective operator
\begin{eqnarray}
    H_e=2d_e
    \left(\begin{array}{cc}
    0 & 0 \\
    0 & \bm{\sigma E} \\
    \end{array}\right)
\label{Wd}
\end{eqnarray}
 is used to express the interaction of $d_e$ with the inner molecular electric
 field $\bm{E}$ ($\bm{\sigma}$ are the Pauli matrices), to avoid
 large numerical cancellation of the terms with opposite signs
 because of Schiff's theorem \cite{Schiff:63, Martensson:92}.  After averaging
 over the electronic coordinates in the molecular wavefunction, one obtains
\begin{eqnarray}
   W_d \Omega = \frac{1}{d_e}
       \langle \Psi_{\Omega}| \sum_i H_e (i)|\Psi_{\Omega} \rangle\ ,
\label{WdOmega}
\end{eqnarray}
 where $ \Psi_{\Omega} $ is wavefunction for the $X^2\Pi_{3/2}$ state.

 To check the accuracy of calculating the wavefunction in the vicinity of the
 iodine nucleus we computed the hyperfine constant $A_\parallel$ (see
 \cite{Dmitriev:92}) and quadrupole coupling constant $eQq_0$,
 where $Q=-710$ millibarn is quadrupole moment of $^{127}$I \cite{Bieron:01},
 $q_0$ is electric field gradient along molecular axis.
 Note, however, that the
  errors in calculated $A_\parallel$, $eQq_0$ and $E_{\rm eff}$
   are not related closely.
 As our recent calculations showed
  \cite{Isaev:04, Petrov:05a},
 the error for $A_\parallel$ presents rather a lower bound estimate for the
 $E_{\rm eff}$ error.
 The quadrupole interaction constant is capable to provide useful information
 about space-inhomogeneous part of the electronic density near a nucleus.
 Unfortunately, the $eQq_0$ value is not a better measure of the calculation
 accuracy of the effective field on the electron than the $A_\parallel$, first
 of all because it doesn't depend directly (like $E_{\rm eff}$ and
 $A_\parallel$) on the electronic spin density near the heavy nucleus.
 In \cite{Chanda:95} the parameters of the Frosch-Foley effective
 spin-rotational Hamiltonian \cite{Frosch:52} were obtained for the ground
 state of HI$^+$.  The connection of the Frosch-Foley parameters $a,b,c$ to
 $A_\parallel$ is
   \footnote{The relation can be obtained by comparing the expression for the
   Frosch-Foley spin-rotational Hamiltonian with the term
   $(I,\hat{A}S)=A_\parallel I_z S_z + A_\perp (I_-S_+ + I_+ S_-)/2$ used in
   \cite{Kozlov:95}.}:

 $$
   A_\parallel = \frac{1}{\Omega}(a \Lambda + (b+c) \Sigma)\ ,
 $$
 where $\Lambda$ is the projection of the
 angular electronic momentum on the molecular axis and $\Sigma{=}S_z$ is
 $z$-projection of the electronic spin.  Accounting for the calculated value of
 $G_\parallel$-factor that is $2.0001$ and $|\Omega|=1.4998$, the ground state
 of HI$^+$ can be reliably classified as $^2\Pi_{3/2}$.

\paragraph*{Methods and calculations.}

 A 25-electron generalized relativistic effective core potential (GRECP)
 \cite{Titov:99} for iodine (its gaussian expansion can be found on our website
 \cite{httpI}) is used at the first step of the two-step calculations of
 HI$^+$, so that the inner shells of iodine, $1s-3d$, are absorbed into the
 GRECP and the $4s$, $4p$, $4d$, $5s$, and $5p$ electrons (as well as an
 electron of hydrogen) are treated explicitly.  Two calculations are carried
 out. In the first one, only seven external electrons of iodine are correlated
 whereas its $4s$, $4p$, $4d$ shells are ``frozen'' within the GRECP approach
 when employing the level-shift technique \cite{Titov:99}.  Thus, a 7-electron
 GRECP version is, in fact, used in the first series of the HI$^+$
 calculations.  In the other calculation, all 25 electrons are explicitly
 correlated.  The terms with the leading configurations $\sigma^2 \pi^3$ are
 calculated where $\sigma$ and $\pi$ are the highest occupied molecular
 orbitals.  The correlation spin-orbit basis sets are optimized in atomic
 two-component relativistic coupled cluster calculations of iodine with single
 and double cluster amplitudes (RCC-SD) using the scheme suggested in
 \cite{Mosyagin:00, Isaev:00}.  As a result, the basis [$5s5p3d2f1g$] was
 generated.
 As our investigation shows removing of $g$-function from the basis set changes
 the RCC-SD results for $A_\parallel$ and $E_{\rm eff}$ on the level of 1\%.
 Thus, contribution from $g$-function to the calculated values can be
 negligible and the basis reduced to [$5s5p3d2f$] without loss of accuracy.
 Such iodine basis was used in 25 electron configuration interaction (CI)
 calculations of HI$^+$.  For hydrogen, the reduced [$4s3p2d$]
 correlation-consistent basis \cite{Dunning:89} was used.

 The HI$^+$ calculations start from a one-component
  closed shell
 SCF computation of the ground state of the neutral HI molecule using the
 spin-averaged GRECP for iodine.  Two-component Fock-space RCC-SD molecular
 calculations or spin-orbit direct CI (SODCI) calculations are then performed.

\paragraph{RCC-SD method:}

 The details on the Fock-space RCC-SD method can be found in Ref.\
 \cite{Kaldor:97, Kaldor:04ba} and references therein. The program package
 {\sc rccsd}
 is used in all RCC calculations mentioned further in
 the article.  The Fock-space RCC calculations start from the ground state of
 HI and use the scheme:
\begin{equation}
 \begin{array}{ccc}
     {\rm HI}&\rightarrow&{\rm HI}^+ \\
 \end{array}
 \label{fockspaceHI}
\end{equation}
 with an electron removed from the $\pi,\pi'$ orbitals.

\paragraph{SODCI method:}

 Spin-orbit direct CI approach with the selected single- and double-excitations
 from some multiconfigurational reference states
 \cite{Buenker:99,Alekseyev:04a} is employed on the sets of different
 {$\Lambda$}S many-electron spin- and space-symmetry adapted basis functions
 (SAFs).  In the {\sc sodci} code,
 the double $C_{2v}$ group, $C_{2v}^*$, is
 used to account for the spin and space symmetry of the HI$^+$ molecule,
 instead of the more restrictive symmetry group $C_{\infty v}^*$, which could,
 in principle, be employed.  In the $C_{2v}^*$ classification scheme, the
 doubly degenerate ground state has the components only in the irreducible
 representation $E^*$.

 The SODCI calculations exploiting relativistic scheme of configuration
 selection \cite{Titov:01} start from some space of the reference functions:
 for 25 correlated electrons 4415 SAFs (see \tref{7e25e}) were included in the
 reference space (``main'' configurations).  These SAFs had the largest
 coefficients in the probing CI calculation.  The single and double excitations
 from this reference space produce about $3\times 10^9$ of SAFs.  Only the most
 important of them, selected by second-order perturbation theory for chosen
 thresholds $T_i$ (see \tref{7e25e}), were included in the subsequent CI
 calculation.  About 1.6, 5.7 and 13 millions of SAFs were selected for
 thresholds $T_1{=}0.01$, $T_2{=}0.001$, $T_3{=}0.0003$, correspondingly.


 Since we are interested in the spin-dependent properties determined mainly by
 the electronic wavefunction near the iodine nucleus, the shape of the valence
 and outer core four-component molecular spinors are restored in the inner core
 of iodine that is done in the paper
  within
 the nonvariational one-center
 restoration scheme (NOCR) (see \cite{Titov:96b,Titov:99,Petrov:02,Titov:05a}
 and references therein).  The RCC calculation of $E_{\rm eff}$  employs the
 finite field method (see Refs.\ \cite{Kunik:71,Monkhorst:77,Petrov:02}).
 In the SODCI calculations conventional approach with the density matrix
 calculation for CI wavefunction was used
 \cite{Petrov:05a}.

\paragraph*{Results and discussion.}

 The results of the RCC and SODCI calculations for 7 and 25 correlated
 electrons of HI$^+$ are presented in \tref{7e25e}. The internuclear distance
     is 3.08 a.u.\
 in accord to the experimental datum \cite{Chanda:95}.


\begin{table}
\caption{
  Calculated $E_{\rm eff}$ (in $\times 10^{24}$~Hz/($e\cdot$cm)), $A_\parallel$
  (in MHz) and quadrupole interaction value $eQq_0$ (in MHz) for the ground
  state $X^2\Pi_{3/2}$ of H$^{127}$I$^+$.  The one-center expansion by $s,p,d$
  spinors within the iodine core is used in the NOCR scheme. Experimental
  values for $A_\parallel$ is 1021 MHz and for quadrupole coupling constant
  $eQq_0$ is $-712.6$ MHz.}

\vspace{1mm}
\begin{ruledtabular}
\begin{tabular}{lrrrrr}

Method   &  & & $E_{\rm eff}$ &$A_\parallel$&$eQq_0$ \\
\hline
\multicolumn{3}{l}{work \cite{Ravaine:05a}
``ionic'' approx. DHF}               & -0.09 &  &  \\
\multicolumn{3}{l}{work \cite{Ravaine:05a}
``covalent'' approx. CI}             & -0.49 &  & \\
\hline
 \multicolumn{6}{c}{\bf AGRECP/SCF/NOCR calculations} \\
 \multicolumn{5}{c}{\it ~~~~~~~~7 electrons } \\
 resticted SCF        & & & 0.008 & 949 & -647 \\
 \multicolumn{5}{c}{\it ~~~~~~~25 electrons } \\
 resticted SCF        & & & 0.010 &1024 & -667 \\
\hline
 \multicolumn{6}{c}{\bf GRECP/RCC/NOCR calculations} \\
\multicolumn{5}{c}{\it ~~~~~~~~7 electrons } \\
RCC-S                          & & & 0.206  &  863& -719\\
RCC-SD                         & & & 0.347  &  881& -708\\
\multicolumn{5}{c}{\it ~~~~~~~25 electrons } \\
           RCC-S                          & & & 0.226  &  906& -807\\
RCC-SD                         & & & 0.345  &  962& -752\\
\hline
 \multicolumn{6}{c}{\bf GRECP/SODCI/NOCR calculations} \\
 Threshold  & SAF number & & & \\
 (mHartree) &            & & & \\
 \multicolumn{5}{c}{\it ~~~~~~~~7 electrons } \\
Mains only &    7\,786 & & 0.294 &  984& -687\\
0.001     &   676\,397 & & 0.335 &  895& -711\\
0.0001    &1\,911\,282 & & 0.336 &  892& -709\\

\multicolumn{5}{c}{\it ~~~~~~~25 electrons } \\
Mains only   & 4\,415  & & 0.333 & 1063& -778\\
0.01     &1\,600\,012  & & 0.299 &  975& -738\\
0.001    &5\,712\,946  & & 0.329 &  971& -743\\
0.0003  &12\,678\,133  & & 0.336 &  968& -745\\
%
\end{tabular}
\end{ruledtabular}
\label{7e25e}
\end{table}

 It should be noted that the authors of paper \cite{Ravaine:05a} considered
 their ``covalent'' result as the final one and presented their ``ionic''
 result only for comparison.  The results of our RCC and SODCI calculations
 give essentially different $E_{\rm eff}$ value than the one obtained in
 \cite{Ravaine:05a} by the configuration interaction calculation for the
 ``covalent'' approximation.  Particularly, the sign of $E_{\rm eff}$ is
 opposite to that by Ravaine et al.  One can see that accounting for
 correlations with the iodine core electrons (occupying the shells $4s$, $4p$
 and $4d$) practically doesn't change the value of $E_{\rm eff}$. The
 importance of accounting for correlations can be seen by comparing the results
 of RCC-S and RCC-SD calculations. In the RCC-S calculations (only with the
 single-body cluster amplitudes) effect of ``spin-polarization'' is taken into
 account analogously to ``unrestricted'' Dirac-Hartree-Fock (DHF) calculation.
 Inclusion of electron correlations in the RCC-SD calculation changes $ E_{\rm
 eff}$ on about 60\%.  At the same time value of $A_\parallel$ is changed only
 on 5\%, that shows that the structure of correlation contributions to
 $A_\parallel$ and $E_{\rm eff}$ is very different.  The same is valid for the
 $eQq_0$ constant, in which correlations contribute less than 10\%.

  The restricted open shell SCF calculations presented in \tref{7e25e}
  were performed with the spin-averaged GRECP (AGRECP) for iodine.
  The value of $E_{\rm eff}$ is more than order of magnitude smaller in
  AGRECP/ SCF/NOCR calculations than in GRECP/RCC-S/ NOCR ones that indicates
  critical importance of accounting for
     one-electron spin-orbit and polarization effects on valence shells
  in calculation of $E_{\rm eff}$.  Similar situation was observed in
  calculations on the a(1) state in PbO \cite{Isaev:04}.  We would like
  to emphasize that after applying the NOCR procedure the proper,
  {\it four-component} shapes of molecular spinors in the core of Pb
  are restored having appropriate
  relativistic behavior
  at the Pb nucleus both after GRECP and AGRECP calculations.
  In the experiment on HI$^+$ in the rotating electric field \cite{Stutz:04}
  the knowledge of the hyperfine coupling value of the proton spin to the
  molecular axis can be usefull.  We calculated the value of $A_\parallel$ on
  the H nucleus in the fremework of above-described AGRECP/SCF/NOCR scheme, the
  value is 0.6 MHz.

   Our results of the SODCI calculation (our final values)
   for the $E_{\rm eff}$, $A_\parallel$ and $eQq_0$  properties are in close
   agreement with the RCC-SD values.
   It means that higher-order cluster amplitudes do not contribute largely
   to these properties.
   Besides,
 outercore-valence correlations practically do not influence on the value of
 $E_{\rm eff}$. On the other hand the value of $A_\parallel$ is increased for
 about 10\% when outercore correlations are taken into account.  It was noticed
 before that rather good accuracy in the calculated $A_\parallel$ value gives
 us
  just
 a lower bound for the
 accuracy of $E_{\rm eff}$.  Taking into account weak dependance of $E_{\rm
 eff}$ from the outercore-valence
 correlations we
  estimate the accuracy of $E_{\rm eff}$ calculation in 10\%.

 In any case our calculations show that the absolute value for $ E_{\rm eff}$
 in $X^2\Pi_{3/2}$ of HI$^+$ is much lower than that in YbF,
 $6.0{\times}10^{24}$~Hz/($e\cdot$cm), and in the metastable $a(1)$ state of
 PbO, $6.1{\times}10^{24}$~Hz/($e\cdot$cm).  Thus, HI$^+$ can be perspective
 candidate for experiments on the EDM search provided that the experimental
 scheme is improved to reach much better statistics or coherence time, than
 that in on-going experiments on YbF and PbO.


\paragraph*{Acknowledgments.}

 The authors are grateful to M.\,Kozlov for drawing our attention to
 the suggested EDM experiment on HI$^+$. This work is supported by the RFBR
 grant 03--03--32335 and, in part, by the CRDF grant RP2--2339--GA--02.
 N.M.\ is also supported by grants of Russian Science Support Foundation
 and the governor of Leningrad district.

\bibliographystyle{./bib/bst/apsrev}
\bibliography{./bib/JournAbbr,./bib/TitovLib,./bib/Titov,./bib/Kaldor,./bib/IsaevLib,./bib/PetrovLib}
\end{document}